\documentstyle[psfig,prl,multicol,aps]{revtex}

\begin{document}

\draft

\title{Self-trapping and stable localized modes in 
nonlinear photonic crystals}

\author{Serge F. Mingaleev$^{1,2}$ and 
Yuri S. Kivshar$^{1}$}

\address{$^1$ Nonlinear Physics Group, 
Research School of Physical Sciences and Engineering \\ 
Australian  National University, Canberra ACT 0200, 
Australia}

\address{$^2$ 
Bogolyubov Institute for Theoretical Physics, 
03143 Kiev, Ukraine}

\maketitle

\begin{abstract} 
We predict the existence of {\em stable} nonlinear localized
modes near the band edge of a two-dimensional reduced-symmetry 
photonic crystal with a Kerr nonlinearity. Employing the 
technique based on the Green function,  we reveal a physical
mechanism of the mode stabilization associated with the 
effective nonlinear dispersion and long-range interaction 
in the photonic crystals.
\end{abstract}

%

\begin{multicols}{2}
\narrowtext


Photonic crystals are usually viewed as an optical analog of 
semiconductors that modify the properties of light similar
to a microscopic atomic lattice that creates a semiconductor 
band-gap for electrons \cite{book}. It is therefore believed 
that by replacing relatively slow electrons with photons as 
the carriers of information, the speed and band-width of 
advanced communication systems will be dramatically 
increased, thus revolutionizing the telecommunication 
industry.  To employ the high-tech potential of photonic 
crystals, it is  crucially important to achieve a dynamical 
tunability of their band gap \cite{john2}. 
This idea can be realized by changing 
the light intensity in the so-called {\em nonlinear photonic 
crystals}, having a periodic modulation of the nonlinear 
refractive index \cite{berger}. Exploration of nonlinear 
properties of photonic band-gap (PBG) materials may 
open {\em new applications of photonic crystals} for 
all-optical signal processing and switching, suggesting an 
effective way to create tunable band-gap structures 
operating  entirely with light. 

One of the important ideas to control all-optical switching 
in the nonlinear regime is to explore the possibility of 
nonlinearity-induced self-trapping and nonlinear localized 
modes in photonic crystals. Existence of nonlinear localized 
modes for the frequencies in the forbidden gaps is usually 
associated with {\em gap solitons}, studied for 
one-dimensional \cite{gaps} 
and  even  two-dimensional (2D) models \cite{john}, 
{\em in the framework of the coupled-mode theory}. 
Validity  of the coupled-mode theory is usually restricted 
by a weak modulation of the refractive index (the so-called 
{\em shallow-grating case}), and therefore {\em this theory 
is not directly applicable to the PBG crystals} where 
modulation of the refractive index is of the order
of its mean value. This  observation calls for a systematic 
analysis of the (still open) problem  of stable 
self-trapping and nonlinear localized modes in PBG 
materials, where the effects of discreteness \cite{mcgurn} and 
long-range interaction \cite{mingaleev} have been recently 
shown to be of crucial importance.

Nonlinear localized modes (also called {\em intrinsic 
localized modes} or 
{\em discrete breathers}) are associated with the energy 
localization that may occur in the absence of any disorder 
and solely due to nonlinearity \cite{modes}. Such nonlinear 
modes can be easily identified as approximate (or sometimes 
exact) analytical solutions of coupled-oscillator 
nonlinear lattice models and also in numerical 
molecular-dynamics simulations, but only very recently 
the first observations of spatially localized nonlinear 
modes have been reported in the physical systems of a very 
different nature \cite{exp_modes}. The main purpose of this 
Letter is to predict the existence of nonlinear localized 
modes, analogous to gap solitons in the continuum limit, 
in 2D nonlinear photonic crystals, and to describe their 
unique properties including stability. 

\begin{figure}
\centerline{\hbox{
\psfig{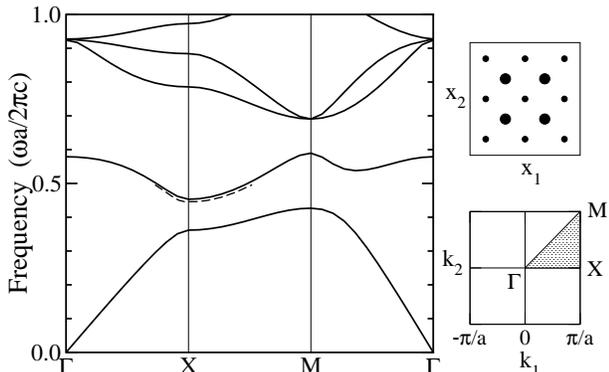}}}
\caption{Band-gap structure of the $E$-polarized 
light in a 2D reduced-symmetry photonic  crystal with 
$r_1 = 0.1a$, $r_2 = 0.05a$, and  $\epsilon = 11.4$.  
Full lines are calculated directly from Eq. (\ref{sys:eq-E-t}) 
by MIT Photonic-Bands program \protect\cite{mpb-prog},  
whereas dashed line is found from linearized Eq. (5).
The top right inset shows a cross-sectional 
view of the photonic crystal and the bottom right inset 
shows the corresponding Brillouin zone.}
\label{fig:band}
\end{figure}

We study nonlinear properties of 2D photonic crystals, 
assuming that their symmetry is reduced by inserting the 
rods made from a {\em Kerr-type nonlinear} 
dielectric material characterized by the third-order
nonlinear susceptibility $\chi^{(3)}$. Specifically, 
we consider a periodic square lattice with the 
lattice spacing $a$ which consists of two types of 
infinitely long cylindrical rods:  the rods of radius $r_1$ 
made from a linear  material and placed at the corners of 
the lattice, and the  rods of radius $r_2$ ($r_2 < r_1$) 
made from a nonlinear  material and placed at the center 
of each unit cell (see  top right inset in 
Fig. \ref{fig:band}). Linear properties of such photonic 
crystals are known \cite{symmetry}. 

We assume that the rods are parallel to the $x_3$ axis, 
so that the system is characterized by the dielectric 
constant $\epsilon(\bbox{x})=\epsilon(x_1, x_2)$. 
In this case the evolution of the $E$-polarized 
(with the electric field $\bbox{E} \,||\, \bbox{x}_3$) light 
propagating in the $(x_1, x_2)$-plane is governed by 
the scalar wave equation
\begin{equation}
\nabla^2 E(\bbox{x}, t) - \frac{1}{c^2} \, \partial_t^2
\left[ \epsilon(\bbox{x}) E \right] = 0 \; ,
\label{sys:eq-E-t}
\end{equation}
where 
$\nabla^2 \equiv \partial_{x_1}^2 + \partial_{x_2}^2$ 
and $E$ is the $x_3$ component of $\bbox{E}$. 
Taking the electric field in the form
$E(\bbox{x}, t) = e^{-i \omega t} \, E(\bbox{x}, t \,|\, 
\omega) \,$, 
where $E(\bbox{x}, t \,|\, \omega)$ is a slowly varying 
envelope, i.e. 
$\partial^2_t E(\bbox{x}, t \,|\, \omega) \ll \omega 
\partial_t E(\bbox{x}, t \,|\, \omega)$, 
Eq. (\ref{sys:eq-E-t}) can be reduced to
\begin{equation}
\left[ \nabla^2  + \epsilon(\bbox{x}) 
\left( \frac{\omega}{c} \right)^2 \right]
E(\bbox{x}, t \,|\, \omega) \simeq -2i 
\epsilon(\bbox{x}) \frac{\omega}{c^2} \, 
\frac{\partial E}{\partial t} \; .
\label{sys:eq-E-omega-t}
\end{equation}
In the stationary case, i.e. when the r.h.s. vanishes, 
Eq. (\ref{sys:eq-E-omega-t}) reduces  to the eigenvalue 
problem which can be solved  in the linear limit 
\cite{book}, i.e. when  
$\epsilon(\bbox{x})$ does not depend on the light intensity. 
In this case,  the frequency spectrum has a band-gap 
structure shown
in  Fig. \ref{fig:band}.  It supports two band gaps, the lower 
of which extends from $\omega=0.426 \times 2 \pi c/a$ to 
$\omega=0.453 \times 2 \pi c/a$. 

A low-intensity light cannot propagate through a photonic 
crystal if the light frequency falls into a band gap. 
However, it has been recently suggested \cite{john} that in 
the case of a 2D periodic medium with a Kerr-type nonlinear 
material, high-intensity light with the frequency inside the 
photonic gap can propagate in the form of {\em finite 
energy solitary waves} ---  {\em 2D gap solitons}. Such 
solitary waves were analyzed in the framework of the 
coupled-mode continuum-limit equations valid for a 
{\em weak modulation} of  the dielectric constant 
$\epsilon(\bbox{x})$. However, in real photonic crystals 
the modulation of $\epsilon(\bbox{x})$ is {\em comparable 
to its average value}, so that the existence and stability 
of localized modes in such structures 
is still an open problem.

\begin{figure}
\centerline{\hbox{
\psfig{figure=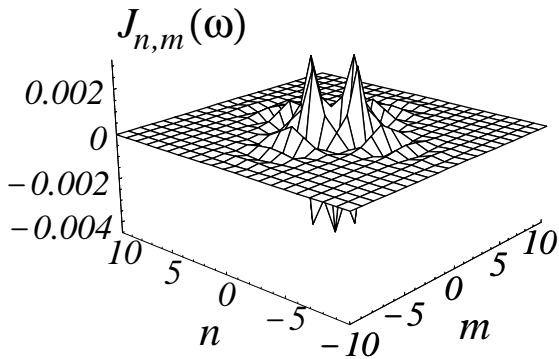,clip=,width=75mm,angle=0}}}
\caption{Coupling coefficients $J_{n,m}(\omega)$ for 
the photonic crystal depicted in Fig. 
\protect\ref{fig:band}. The frequency 
$\omega=0.4456 \times 2\pi c/a$ is in the lower band gap.
The contribution from $J_{0,0}=0.039$ is not shown.}
\label{fig:Jnm}
\end{figure}

More specifically, the coupled-mode equations are valid if 
and only if the band gap $\Delta \omega^2$ is vanishingly 
small, i. e. $\Delta \omega^2  \sim A^2$ where $A$ is an 
effective amplitude of the mode that is treated 
in the multi-scale asymptotic expansions \cite{kiv} 
as a small parameter. If we apply this
model to describe nonlinear modes in a wider gap (see, e.g.,
discussions in Ref. \cite{kiv}), we obtain a 2D cubic nonlinear
Schr{\"o}dinger (NLS) equation 
known to possess {\em no stable localized solutions}. 
Moreover, 2D localized modes of the coupled-mode equations 
are expected to possess
{\em an oscillatory instability} recently discovered 
for a broad class of 1D coupled-mode
equations \cite{dima}. Thus, it is clear that, if nonlinear 
localized modes do exist in realistic PBG materials, their 
existence and stability should be associated with 
{\em different physical mechanisms} not accounted for by 
simplified continuum coupled-mode models.


To study the nonlinear modes in such structures, 
we consider the nonlinear rods of small radius $r_2$ as 
``defects'' embedded into the linear photonic crystal 
formed by a square lattice of the rods of larger radius 
$r_1$ ("diatomic crystal"). Then, writing the dielectric 
constant 
$\epsilon(\bbox{x})$ as a sum of two periodic terms, 
$\epsilon(\bbox{x})=\epsilon_{1}(\bbox{x}) + 
\epsilon_{2}(\bbox{x}\,|\,E) \, ,$ 
where $\epsilon_{1}(\bbox{x})$ describes the linear photonic 
crystal and $\epsilon_{2}(\bbox{x}\,|\,E)$ corresponds to a 
lattice of  nonlinear defect rods, one can present 
Eq. (\ref{sys:eq-E-omega-t}) in 
the form \cite{mcgurn,mingaleev} 
\begin{equation}
E(\bbox{x}, t \,|\, \omega) = 
\int d^2\bbox{y} \,\,\, G(\bbox{x}, 
\bbox{y} \,|\, \omega) \, \hat{\cal L} \, 
E(\bbox{y}, t \,|\, \omega) \; ,
\label{sys:eq-green-int}
\end{equation}
where we introduce the Green function 
$G(\bbox{x}, \bbox{y} \,|\, \omega)$ 
of the linear photonic crystal (see, e.g., 
Ref. \cite{Maradudin:1993:PBGL} for its properties)
and the linear operator 
\begin{eqnarray}
\hat{\cal L} = \left( \frac{\omega}{c} \right)^2 
\epsilon_{2}(\bbox{x}\,|\,E) 
+ 2 \, i \, \epsilon(\bbox{x}) 
\frac{\omega}{c^2} \frac{\partial}{\partial t} \; . 
\end{eqnarray} 
Now, using the indices $n$ and $m$ for numbering the 
nonlinear rods in the $x_1$ and $x_2$ directions,
 we can describe their positions by the vectors 
$\bbox{x}_{n,m} = n \, \bbox{a}_{1} + m \, \bbox{a}_{2}$, 
where $\bbox{a}_1$ and $\bbox{a}_2$
are the primitive lattice vectors of the 2D photonic
crystal, and write 
\begin{displaymath}
\epsilon_2(\bbox{x}\,|\,E) = \left\{\epsilon_{2}^{(0)} +
\chi^{(3)} |E(\bbox{x}, t \,|\, \omega)|^2\right\} 
\sum_{n,m} \theta (\bbox{x}-\bbox{x}_{n,m}) \; ,
\end{displaymath}
where 
$\theta (\bbox{x}) = 1$ for $|\bbox{x}| \leq r_2$ 
and $\theta (\bbox{x}) = 0$ otherwise.
The parameter $\epsilon_{2}^{(0)}$ 
is the dielectric constant of the
defect rods in the linear limit, while the term 
$\chi^{(3)} |E|^2$ 
takes into account a contribution due to 
the Kerr nonlinearity. 
If the radius $r_{2}$ of the defect rods is 
sufficiently small, the electric field 
$E(\bbox{x}, t \,|\, \omega)$ inside them 
is almost constant,  and Eq. (\ref{sys:eq-green-int}) can be 
approximated \cite{mcgurn,mingaleev} by the 
{\em discrete nonlinear equation} 
\begin{eqnarray}
\label{sys:eq-E-disc}
i \sigma \frac{\partial}{\partial t} E_{n,m} 
- E_{n,m} \nonumber \\ 
+ \sum_{k,l} J_{n-k, \, m-l}(\omega) 
\left\{ \epsilon_{2}^{(0)} + \chi^{(3)} |E_{k,l}|^2
\right\} E_{k,l} = 0\; ,
\end{eqnarray}
for the amplitudes of 
the electric field $E_{n,m}(t \,|\, \omega) \equiv 
E(\bbox{x}_{n,m}, t \,|\, \omega)$ calculated at the 
defect rods. The parameter $\sigma$ and 
the coupling coefficients 
\begin{equation}
J_{n,m}(\omega) = \left( \frac{\omega}{c} \right)^2
\int_{r_2} d^2 \bbox{y} \,\,\,
G(\bbox{x}_{0,0}, \bbox{x}_{n,m} + \bbox{y} \,|\, \omega ) 
\label{sys:Jn}
\end{equation}
are determined by the Green function 
$G(\bbox{x}, \bbox{y} \,|\, \omega)$ which we calculate
numerically by the FDTD method 
\cite{Ward:1998:PRB} with the spatial step $\Delta x=0.01$
to $0.03$ and time step $\Delta t=0.005$, for the lattice 
 $24a \times 24a$. The Green 
function and, therefore,  the coupling coefficients 
$J_{n,m}(\omega)$ in 2D photonic crystals are 
usually long-ranged functions.  
For instance, for the case of Fig. 
\ref{fig:Jnm} we obtain $J_{n,0} \simeq 0.012 \, (-1)^n 
\, \exp(-0.66\,|n|)$ for $n \geq 2$, so that one should 
take into account the interaction between at least 10 
neighbors to achieve a good accuracy for the spectrum. 
By this means, 
Eq. (\ref{sys:eq-E-disc}) is {\em a nontrivial 
long-range generalization} of the 2D discrete NLS 
equations extensively studied during the last 
decade \cite{DNLS}. 
We have checked the accuracy of 
the approximation provided by Eq. (\ref{sys:eq-E-disc}) 
by solving it in the linear limit. 
The low-frequency part of this dependence for 
$\epsilon_2^{(0)}=11.4$ is depicted in Fig. 
\ref{fig:band} by a dashed line; it has a minimum at 
$\omega=0.446 \times 2\pi c/a$ being in a good 
agreement with the band edge calculated directly from Eq. 
(\ref{sys:eq-E-omega-t}). This lends a support to the validity 
of Eq. (\ref{sys:eq-E-disc}) and allows us to use this discrete model for 
studying nonlinear properties. 


\begin{figure}
\centerline{\hspace{-30mm}\hbox{
\psfig{figure=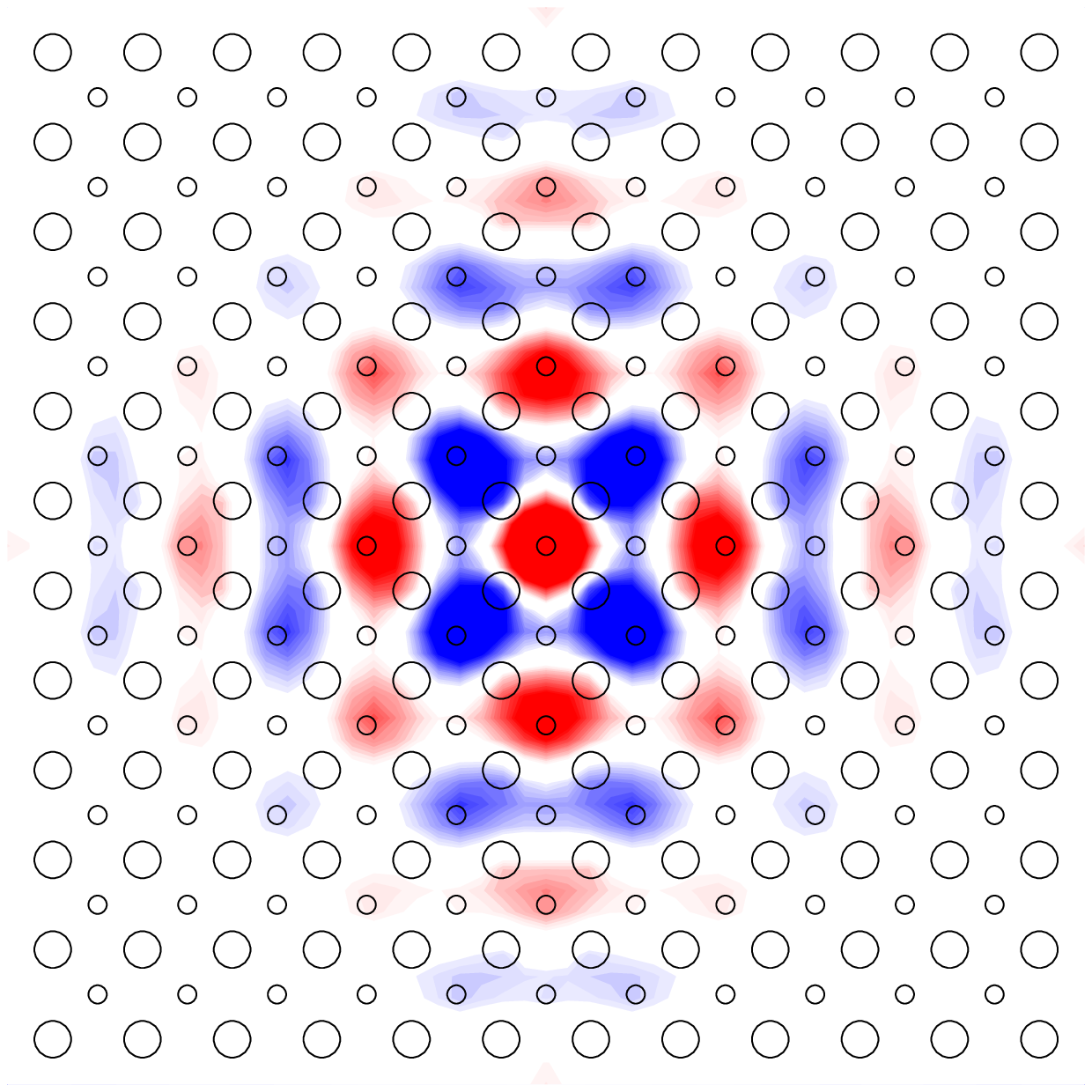,clip=,width=33mm,angle=0}}}
\vspace{-14mm}
\centerline{\hbox{
\psfig{figure=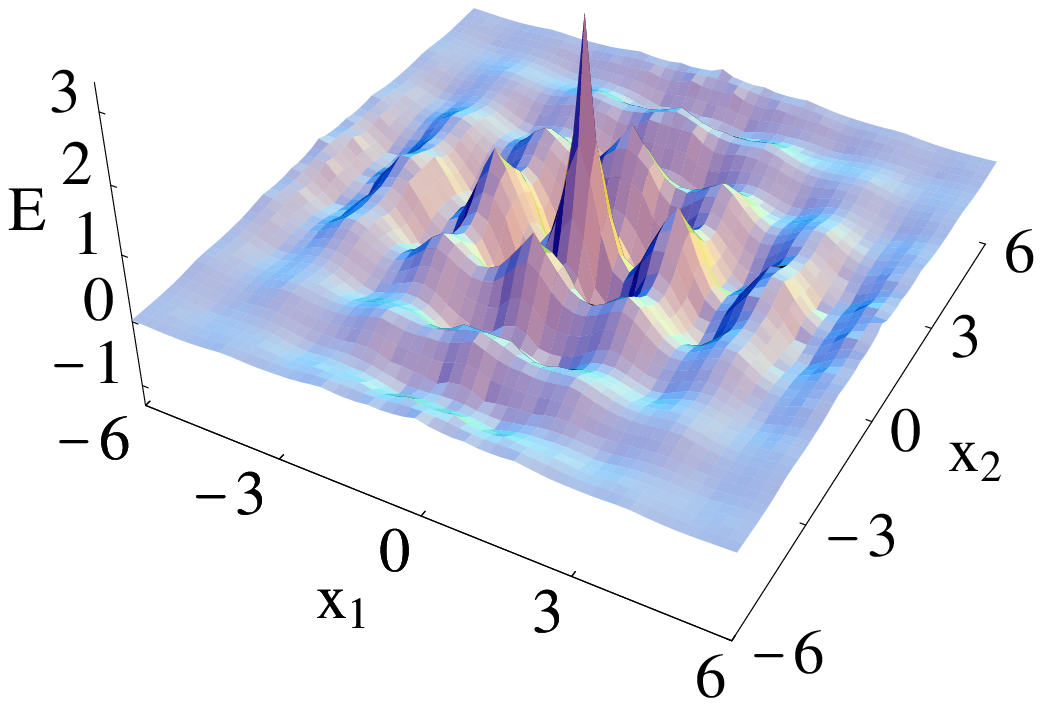,clip=,width=65mm,angle=0}}}
\caption{(Color) Top and 3D views of a nonlinear localized 
mode in the 2D photonic crystal shown in Fig. \protect\ref{fig:band}
($\omega=0.442 \times 2\pi c/a$).}
\label{fig:mode}
\end{figure}

Stationary nonlinear localized modes have been calculated 
numerically by the Newton-Raphson iteration scheme 
on the lattice $170a \times 170a$. 
Assuming that the linear rods are made from GaAs whereas 
the nonlinear rods are made from some nonlinear material 
(which we do not specify varying $\epsilon_2^{(0)}$),  we find
{\em a continuous family of nonlinear localized 
modes}, and a typical example [smoothed by continuous 
optimization for Eq. (\ref{sys:eq-green-int})] of such a mode 
is shown in Fig. \ref{fig:mode}. 
At first glance,  this mode can be regarded as a donor state 
created by a single defect rod with larger dielectric 
constant.  However, in the nonlinear case the mode 
stability becomes a critical issue. It can be determined 
from the so-called Vakhitov-Kolokolov stability criterion
extended to the 2D discrete NLS models in Ref. \cite{Laedke}.
According to this criterion, the nonlinear localized states 
with $dQ/d\omega < 0$ are stable, and they are unstable 
otherwise. Here 
\begin{equation}
Q(\omega) = \sum_{n,m} |E_{n,m}|^2 \; ,
\label{sys:norm}
\end{equation}
is the conserved mode power proportional to the 
energy of the electric field  accumulated in the nonlinear 
localized mode. 

\begin{figure}
\centerline{\hbox{
\psfig{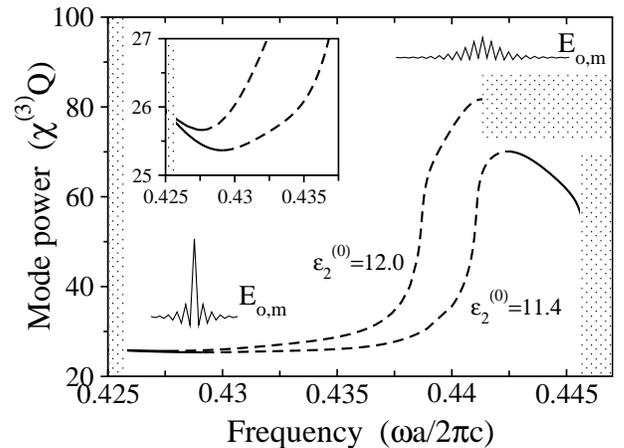}}}
\caption{Power $Q$  vs. frequency $\omega$ for 
the 2D nonlinear localized modes in the photonic crystal 
of Fig. \protect\ref{fig:band} with two different 
$\epsilon_2^{(0)}$. Solid lines -- stable modes, 
dashed lines -- unstable modes. 
Grey areas show the projected band structure of the 
crystal. Insets show typical profiles of stable modes,
and an enlarged part of the power dependence.} 
\label{fig:norm}
\end{figure}

As is well known (see, e.g., Refs. \cite{Laedke}) 
in the 2D discrete cubic NLS equation, only high-amplitude 
localized modes are stable, whereas no stable modes exist in
the continuum limit. For our model, the high-amplitude modes
are also stable (see inset in Fig. \ref{fig:norm}), 
but they are not accessible under realistic conditions: 
To excite such modes one  should increase the refractive index
at the mode center in more than 2 times. 
Thus, for realistic conditions and relatively small values of 
$\chi^{(3)}$, only low-amplitude localized modes become a 
subject of much interest since they can be excited in 
experiment. However, such modes in unbounded 2D NLS models 
are always unstable and either collapse or spread out 
\cite{DNLS}.  Here we reveal that, in a sharp contrast to the 
2D discrete NLS models discussed earlier in various 
applications, the low-amplitude localized modes 
of Eq. (\ref{sys:eq-E-disc}) can be stabilized 
due to {\em nonlinear long-range dispersion} inherent to 
the photonic crystals.  It should be emphasized that such 
stabilization does not occur in the models with only {\em linear long-range} 
dispersion \cite{DNLS}. 

In order to gain a better insight into the stabilization 
mechanism, we have carried out the studies of Eq. 
(\ref{sys:eq-E-disc}) for the exponentially decaying 
coupling coefficients $J_{n,m}$. Our results show that 
the most important factor which determines stability 
of the low-amplitude localized modes is a ratio of the 
coefficients 
at the local nonlinearity ($\sim J_{0,0}$) and the 
nonlinear dispersion ($\sim J_{0,1}$). If the coupling 
coefficients $J_{n,m}$ decrease with the distances $n$ and $m$ 
rapidly, the low-amplitude modes of Eq. (\ref{sys:eq-E-disc}) 
with $\epsilon_2^{(0)}=11.4$ are essentially stable 
for $J_{0,0}/J_{0,1} \lesssim 13$. However, this estimation 
is usually lowered because the stabilization is favored 
by the presence of long-range interactions. 

It should be mentioned that the stabilization of 
low-amplitude 2D localized modes is not inherent to all types 
of nonlinear photonic crystals. On the contrary, the photonic 
crystals must be {\em carefully designed} to support 
{\em stable low-amplitude nonlinear modes}. For example, in the
photonic crystal considered above such modes are stable 
at least for $11 < \epsilon_2^{(0)} < 12$, however they become 
unstable for $\epsilon_2^{(0)} \geq 12$ 
(see Fig. \ref{fig:norm}). 
The stability of these modes can also be controlled by 
varying $r_2$, $r_1$, or $\epsilon_1$. 
Thus, experimental observation of the nonlinear localized 
modes would require not only the use of photonic materials 
with a relatively large nonlinear refractive index (such as 
GaAs/AlAs periodic structures \cite{algas} or polymer PBG 
crystals \cite{jap}), but also a fine adjustment of the
parameters of the photonic crystal. The latter can be 
achieved, in principle, by employing the surface coupling 
technique \cite{ast} that is able to provide 
coupling to  specific points of the dispersion curve, 
opening up a very straightforward way to access nonlinear 
effects.

In conclusion, we have developed a consistent theory of 
nonlinearity-induced self-trapping effects in 2D nonlinear 
photonic crystals  and predicted the possibility of the 
energy localization near the bandgap edge in the form 
of stable 2D nonlinear localized modes.


Yuri Kivshar thanks V. Astratov, K. Busch,  S. John,
A. McGurn, M. Scalora, and C. Soukoulis for encouraging 
comments and discussions. 
Serge Mingaleev is indebted to Yu.B. Gaididei for 
useful remarks. The work has been partially supported by 
the Australian Research Council and the Performance and 
Planning Foundation grant of the Institute of Advanced Studies.



\end{multicols}

\begin{thebibliography}{99}

\bibitem{book} J. D. Joannoupoulos, R. B. Meade, and 
J. N. Winn, {\em Photonic Crystals} 
(Princeton University Press, 1995).

\bibitem{john2} See, e.g.,  K. Busch and S. John, 
Phys. Rev. Lett. {\bf 83}, 967 (1999), 
and discussions therein.

\bibitem{berger} See, e.g., M. Scalora {\em et al.}, 
Phys. Rev. Lett. {\bf 73}, 1368 (1994); 
P. Tran, Phys. Rev. B {\bf 52}, 10673 (1995).

\bibitem{gaps} W. Chen and D. L. Mills, 
Phys. Rev. Lett. {\bf 58}, 160 (1987); D. N. Christodoulides 
and R. I. Joseph, Phys. Rev. Lett. {\bf 62}, 1746 (1989); 
see also the review paper: C. M. de Sterke and J. E. Sipe, 
in {\em Progress in Optics XXXIII}, edited by E. Wolf 
(Elsevier, Amsterdam, 1994), p. 203.

\bibitem{john} S. John and N. Ak\"ozbek, 
Phys. Rev. Lett. {\bf 71}, 1168 (1993); 
Phys. Rev. E {\bf 57}, 2287 (1998).

\bibitem{mcgurn} A. R. McGurn, Phys. Lett. A 
{\bf 260}, 314 (1999).

\bibitem{mingaleev} S. F. Mingaleev {\em et al.},
Phys. Rev. E {\bf 62}, 5777 (2000).

\bibitem{modes} See, e.g., A. J. Sievers and 
S. Takeno, Phys. Rev. Lett. {\bf 61}, 970 (1988); 
S. Flach and C. R. Willis,  Phys. Rep. {\bf 295}, 181 (1998).

\bibitem{exp_modes} 
H. S. Eisenberg {\em et al.}, 
Phys. Rev. Lett. {\bf 81}, 3383 (1998); 
B. I. Swanson {\em et al.}, 
Phys. Rev. Lett. {\bf 82}, 3288 (1999); 
U. T. Schwarz {\em et al.}, 
Phys. Rev. Lett. {\bf 83}, 223 (1999); 
E. Trias {\em et al.}, Phys. Rev. Lett. {\bf 84}, 741 (2000); 
P. Binder {\em et al.}, Phys. Rev. Lett. {\bf 84}, 745 (2000).

\bibitem{symmetry} C. M. Anderson and K. P. Giapis, 
Phys. Rev. Lett. {\bf 77}, 2949 (1996); 
Phys. Rev. B {\bf 56}, 7313 (1997).

\bibitem{mpb-prog} 
S. G. Johnson, http://ab-initio.mit.edu/mpb/ 

\bibitem{kiv} Yu. S. Kivshar {\em et al.}, 
Int. J. Mod. Phys. B {\bf 9}, 2963 (1995).

\bibitem{dima}
I. V. Barashenkov {\em et al.},  
Phys. Rev. Lett. {\bf 80}, 5117 (1998); 
A. De Rossi {\em et al.},  
Phys. Rev. Lett. {\bf 81}, 85 (1998).

\bibitem{Maradudin:1993:PBGL}
A.~A. Maradudin and A.~R. McGurn,  in 
{\em Photonic Band Gaps and Localization}, 
{\em NATO ASI Series B} {\bf 308},
Ed. C.~M. Soukoulis (Plenum Press, New York, 1993), p. 247.

\bibitem{Ward:1998:PRB}
A.~J. Ward and J.~B. Pendry, 
Phys. Rev. B {\bf 58}, 7252  (1998).

\bibitem{DNLS} See, e.g., 
V.~K. Mezentsev {\em et al.}, JETP Lett. {\bf 60}, 829 (1994); 
S. Flach {\em et al.}, Phys. Rev. Lett. {\bf 78}, 1207 (1997);
P.~L. Christiansen {\em et al.}, 
Phys. Rev. B {\bf 57}, 11303 (1998).

\bibitem{Laedke}
E.~W.~Laedke {\em et al.}, JETP Lett. {\bf 62}, 677 (1995);
Phys. Rev. E {\bf 52}, 5549 (1995); 
Yu. B. Gaididei {\em et al.}, Phys. Rev. B {\bf 55}, 
R13365 (1997).

\bibitem{algas} 
P. Millar {\em et al.}, Opt. Lett. {\bf 24}, 685 (1999); 
A.A. Helmy {\em et al.}, Opt. Lett. {\bf 25}, 1370 (2000).

\bibitem{jap} 
S. Shoji and S. Kawata,  
Appl. Phys. Lett. {\bf 76}, 2668 (2000).

\bibitem{ast} 
V. N. Astratov {\em et al.}, 
Phys. Rev.  B {\bf 60}, R16255 (1999).

\end{thebibliography}
\end{document}